# The fundamental scales of structures from first principles

Scott Funkhouser
Department of Physics, The Citadel
171 Moultrie St., Charleston, SC, 29409

ABSTRACT

Five fundamental scales of mass follow from holographic limitations, a self-similar law for angular momentum and the basic scaling laws for a fractal universe with dimension 2. The five scales correspond to the observable universe, clusters, galaxies, stars and the nucleon. The fundamental scales form naturally a self-similar hierarchy, generating new relationships among the parameters of the nucleon, the cosmological constant and the Planck scale. There is implied a sixth fundamental scale that corresponds to the electrostatic force within an atom. Identifying the implied scale as such leads to new relationships among the fundamental charge, the mass of the electron and cosmological parameters. These considerations also suggest that structures on the scale of galaxies and larger must be bound by non-Newtonian forces.

*1. Introduction*

The combination of three well-established components of modern physics leads to a set of critical scales for the masses of astronomical bodies and particles. Those critical scales correspond to the observable cosmos, clusters of galaxies, galaxies, solar systems and the nucleon. The three components that generate those fundamental scales are the physics of a fractal universe, a self-similar scaling law for angular momentum and holographic limits on information. In Sections 2, 3 and 4 of this work, basic relationships from those physical components are presented. In Sections 5 – 9 the fundamental structural scales are obtained by requiring consistency among those relationships.

In Section 10 the critical scales of mass derived in Sections 5 – 9 are shown to form naturally a self-similar hierarchy based on powers of a pure number that is a function of fundamental constants. The parameters of the nucleon and the cosmological constant are accordingly related to the Planck scale. In Section 11 it is shown that the most natural identification of the one missing term in the hierarchy would relate the fundamental unit of charge and the mass of the electron to the other fundamental constants.

*2. Laws of fractal structure*

The universe exhibits compelling indications that it is arranged, on the predominant structural levels, according to fractal scaling laws [1],[2]. Any given structural level in a fractal universe consists primarily of an ensemble of roughly identical bodies, each of which is contained within a characteristic spherical cell. According to the basic properties of fractal structures, the characteristic mass $M_x$ of bodies on some structural level $x$ and the associated cell-radius $R_x$ are related according to

$$M_x = qR_x^D, \quad (2.1)$$

where $q$ is a real constant. Associated with each fractal body on the structural level $x$ is a characteristic quantity of action $A_x$, which is related to the mass of the body according to

$$A_x \sim sM_x^{(D+1)/D}, \quad (2.2)$$

where $s$ is a constant [2].

There is considerable theoretical and empirical justification for expecting that the fractal dimension $D$ representing the predominant structural levels of the cosmos is 2 [1], [2]. Although the universe features certain structures that are best described by a fractal dimension that is different than 2 [3], the remainder of this work addresses those predominant features that are well represented by $D=2$.

Given the fractal dimension, the constants $q$ and $s$ may be determined accordingly. With $D=2$ the constant $q$ is, conveniently, $g_0/G$, where $g_0$ is the characteristic gravitational field $GM_x/R_x^2$ of the bodies on any fractal structural level and $G$ is the Newtonian gravitational coupling. The gravitational fields of solar systems, galaxies, clusters and the observable universe (in this epoch) are all of order near $10^{-10}$m/s$^2$, which is a compelling validation that $D=2$ on the predominant structural levels. Thus, $q$ should be of order near $10^0$kg/m$^2$.

If the nucleon is part of the fractal structure of the cosmos then $q$ must also apply to the parameters of the nucleon. The characteristic cell-radius of the nucleon may be approximated by the Compton wavelength $l_n = \hbar/(m_n c)$, where $\hbar$ is the Planck quantum, $m_n$ is the nucleon mass and $c$ is the vacuum-speed of light. The constant $q$ should be therefore, roughly,

$$q \sim \frac{m_n}{l_n^2} \sim \frac{m_n^3 c^2}{\hbar^2}, \qquad (2.3)$$

which is of order near $10^2$kg/m$^2$. Note that both methods of determining $q$ are only approximate and are likely to differ from the actual value by an order of magnitude. Thus, both estimates of $q$ are consistent with a value of the order $10^1$kg/m$^2$.

The constant $s$ in (2.2) may be obtained by similar methods. The characteristic action of a virialized body, such as a galaxy, is well represented by the product of the characteristic energy and the characteristic relaxation time [2]. For a virialized body with a mass $M$ and characteristic radius $R$, the characteristic energy is roughly $GM^2/R$. The characteristic relaxation time of a virialized body is roughly its rotation period $R/v$, where $v$ is the orbital velocity $(GM/R)^{1/2}$. The action is thus roughly $(GM^3 R)^{1/2}$, which is also roughly the characteristic angular momentum of the body. For a typical galaxy whose mass is of the order $10^{42}$kg and whose characteristic radius is of the order $10^{20}$m the action is roughly $10^{69}$Js. Thus the constant $s$ should be of the order $10^{69}$Js/$(10^{42}$kg$)^{3/2} \sim 10^5$Js/kg$^{3/2}$.

The constant $s$ may also be determined from the parameters of the nucleon. The characteristic action of a nucleon is, roughly, the Planck quantum $\hbar$. It follows, therefore, from applying (2.2) to the nucleon that

$$s \sim \frac{\hbar}{m_n^{3/2}}, \qquad (2.4)$$

which is of the order $10^6$Js/kg$^{3/2}$ and, given the associated uncertainties, consistent with the value for $s$ obtained from galactic parameters. With $q$ given by (2.3), the constant $s$ in (2.4) may also be expressed as

$$s \sim \frac{c}{q^{1/2}}. \qquad (2.5)$$

The expressions for $s$ and $q$ in terms of fundamental constants lead to a convenient and physically suggestive expression for the action in (2.2). With the expression for $s$ in (2.5), the fractal scaling law for action in (2.2) becomes, for any given fractal level $x$,

$$A_x \sim \frac{c}{q^{1/2}} M_x^{3/2}. \tag{2.6}$$

With a substitution from (2.1), the relationship in (2.6) leads to

$$A_x \sim M_x c R_x. \tag{2.7}$$

The relationship in (2.7) suggests a connection between astronomical scales and particle physics, since any particle of mass $m$ obeys a similar relationship, $\hbar = mcl$, where $l$ is the Compton wavelength of the particle.

*3. A universal scaling law for angular momentum*

There exists a universal scaling law for the angular momentum of astronomical bodies that is both theoretically and empirically supported. The rotational angular momentum $J$ of any given astronomical body is well represented by

$$J \sim pM^2, \tag{3.1}$$

where $M$ is the mass of the body and $p$ is, presumably, a new constant of nature [4],[5]. The scaling law in (3.1) follows from principles of self-similarity [4], and the existence of the new fundamental constant $p$ is justified by considerations of the unification of gravitation and particle physics [5]. The relationship in (3.1) has been validated on scales ranging from asteroids to large-scale structures and the constant $p$ has been determined empirically to be roughly $8 \times 10^{-17} m^2 kg^{-1} s^{-1}$ [5]. The constant $p$ has the same units as, and is apparently about two orders of magnitude larger than, $G/c$.

The angular momentum given by the scaling law in (3.1) may be compared to the ordinary, mechanical expression for the angular momentum of a body. Consider an astronomical body of mass $M$ and characteristic radius $a$, that is rotating with a characteristic angular frequency $\omega$. The rotational angular momentum of the body is roughly $Ma^2\omega$, and equating that term to the term in (3.1) leads to an expression for the characteristic radius $a$,

$$a \sim \left(\frac{pM}{\omega}\right)^{1/2}. \tag{3.2}$$

If the body is virialized then the rotation frequency is roughly $(GM/a^3)^{1/2}$. The characteristic radius $a$ that corresponds to the virialized rotation is, according to (3.2),

$$a \sim \frac{p^2 M}{G} \tag{3.3}$$

The virialized radius in (3.3) is also the minimum possible radius of a gravitationally bound body whose angular momentum is given by (3.1). That is because the virialized rotation frequency is roughly the maximum possible rotation frequency of any gravitationally bound body, since the rotation frequency $(GM/a^3)^{1/2}$ corresponds to the frequency at which the orbital velocity of the outermost elements is of the order the escape velocity of the system.

*4. Holographic limitations*

According to holographic principles and the thermodynamic models of black holes the maximum number $N(R)$ of bits of information that may be registered by any sphere of radius $R$ is roughly

$$N(R) \sim \frac{R^2}{l_P^2}, \tag{4.1}$$

where $l_P$ is the Planck length [6]. The number of bits required to represent some quantity $L$ of action or angular momentum is roughly $L/\hbar$. Therefore, the maximum possible quantity $S(R)$ of action or angular momentum that could be registered by any sphere of radius $R$ is approximately

$$S(R) \sim \hbar \frac{R^2}{l_P^2}. \tag{4.2}$$

There is another important holographic limit that is relevant for astronomical bodies. The angular momentum $J$ associated with a sphere of radius $R$ containing a mass $M$ is limited according to [7]

$$J < McR. \tag{4.3}$$

The largest mass that could be contained within a sphere of radius $R$ is roughly $Rc^2/G$. Thus, the largest possible upper bound that could be given by (4.3) is roughly $Rc^3/G$, which is identical to (4.2).

These considerations lead to a significant physical conclusion about a fractal universe. The action of a body on a fractal structural level is, according to (2.7), roughly the maximum possible angular momentum that could be associated with the mass and cell-radius of the body. Since the action of a gravitationally virialized body is roughly equal to its angular momentum, the angular momentum of gravitationally virialized bodies constituting a fractal structural level with dimension $D=2$ must be of order near the maximum possible angular momentum allowed by the holographic bound.

5. *The cosmic mass*

The combination of the basic physical principles outlined in Sections 2 – 4 leads to a set of five critical scales of mass. The five scales of mass are expressed in terms of fundamental constants including the parameters $p$ and $q$ associated with a self-similar universe. Since the parameters $q$ and $p$ are not well specified, there is some uncertainty associated with the numerical values of the critical masses. Furthermore, the five critical scales are derived here based on idealized, spherical representations of the basic structures of the universe. Given those sources of error and uncertainty the scales derived here may differ from the structures to which they correspond by an order of magnitude or two. However, such discrepancies are small in comparison to the scales addressed by this work.

The first critical mass obtained here follows from considering that the holographic bound imposes a limit on the action that could be associated with a system. Any body that is part of the fractal structure of the universe should have an action given by (2.7). However, the limit in (4.2) establishes an upper bound on the action of a body. Let the mass $M_u$ and associated cell-radius $R_u$, given by (2.1), be defined so that they satisfy the critical condition generated by (2.7) and (4.2), being

$$M_u c R_u \sim \hbar \left( \frac{R_u}{l_P} \right)^2. \tag{5.1}$$

Eq. (5.1) leads to

$$M_u \sim \frac{c^4}{G^2 q}, \tag{5.2}$$

which is of the order $10^{53}$kg. The corresponding cell-radius $R_u$ is $c^2/(Gq)$, which is of the order $10^{26}$m. Those parameters correspond roughly to the mass and horizon of the observable universe in this current epoch.

Note that, according to the analysis in Section 4, the only body that could satisfy (5.1) is a body whose mass $m$ and characteristic radius $r$ satisfy the Schwarzschild condition $Gm/r \sim c^2$. It follows from the standard model of cosmology that the observable mass and particle horizon of the universe satisfy that condition during the era of matter-dominance and still roughly at this time, which is apparently near the beginning of the era of vacuum-dominance. However, in previous ages the observable mass of the universe was smaller than $M_u$. It is remarkable that the mass $M_u$ is of order near the observable mass only when the universe is at least as old as its current age. That coincidence becomes even more remarkable given that the current observable mass is special also because of the putative cosmological constant. These considerations are addressed in Section 10, leading to new relationships among the fundamental parameters of nature.

*6. The scale of the largest structures*

A second critical scale of mass follows from considering the holographic limit on the angular momentum of a system. Let the mass $M_c$ and radius $R_c$ be defined such that they satisfy the critical condition obtained by equating the angular momentum in (3.1) to the maximum possible angular momentum given by (4.3),

$$pM_c^2 \sim M_c c R_c. \qquad (6.1)$$

The mass $M_c$ that satisfies (6.1) and (2.1) is

$$M_c \sim \frac{c^2}{p^2 q}, \qquad (6.2)$$

which is of order near $10^{47}$kg. From (2.1), the fractal cell-radius associated with the maximum mass in (6.2) is $c/(pq)$, which is of the order $10^{23}$m. These parameters correspond roughly to the largest clusters of galaxies.

*7. The galactic scale*

A third fundamental scale of mass follows from considering the point at which the virialized radius in (3.3) becomes equal to the radius of the fractal cell. Let the mass $M_g$ and cell-radius $R_g$ be defined such that they satisfy the critical condition

$$\frac{p^2 M_g}{G} \sim R_g. \qquad (7.1)$$

With (2.1), the condition in (7.1) leads to

$$M_g \sim \frac{G^2}{p^4 q}. \qquad (7.2)$$

The mass in (7.2) is roughly $10^{43}$kg, which is of order near the scale of galactic mass. The characteristic cell-radius $R_g$ associated with the mass in (7.2) is $G/(pq)$, which is of the order $10^{21}$m, corresponding roughly to the characteristic galactic radius.

The critical mass in (7.2) is particularly significant since the virialized radius on the left side of (7.1) is also the minimum possible radius for a body that is bound by Newtonian gravitation. For any fractal body whose mass is greater than some mass near $M_g$ the minimum Newtonian radius of the body would be larger than the fractal cell-radius given by (2.1). However, in order to be consistent with the basic requirements of fractal structure, the body must nonetheless be confined within its respective cell. Thus, there must be some non-Newtonian force acting to confine that body within the fractal cell. The required force must be stronger than Newtonian gravitation since the force must confine the body to a sphere whose radius is smaller than the radius of the sphere to which Newtonian gravity could confine the sphere. This conclusion suggests that the well-known problem with modeling the dynamics of galaxies and clusters with Newtonian physics is due to the necessarily non-Newtonian nature of the forces associated with fractal structures. It is important to note that, according to this analysis, the existence of some invisible matter (i.e. dark matter) would not solve the dynamical problem since non-Newtonian forces would be required as long as the mass were greater than the critical mass $M_g$. ($M_g \sim 10^{43}$kg is greater than the typical galactic mass, which is of the order $10^{42}$kg. However, such a discrepancy is easily attributed to uncertainties in the terms $p$ and $q$ and the fact that the derivation of $M_g$ was based on an idealized body that is expected to misrepresent actual galaxies.)

*8. The scale of stars*
Let the preponderance of the mass associated with some fractal structural level be concentrated by gravity into a body whose characteristic radius is significantly smaller than the cell-radius. (For example, the mass of the structural level associated with a solar system is concentrated into at least one central star whose characteristic radius is much smaller than the associated cell-radius.) Since the body must have an action given by (2.7), a critical condition is established by the point at which the action associated with the body is of order the maximum action allowed to the body according to the holographic bound in (4.2). Let the mass $M_s$, cell-radius $R_s$ and characteristic radius $a_s$ be defined such that they satisfy that critical condition, which is

$$M_s c R_s \sim \hbar \frac{a_s^2}{l_P^2}. \tag{8.1}$$

(Note that the condition in (8.1) differs from the condition in (5.1) in that the limit in (8.1) is determined by the characteristic radius of the body, not by the cell-radius.) With a substitution from (2.1), (8.1) leads to

$$a_s \sim \left(\frac{GM_s^{3/2}}{c^2 q^{1/2}}\right)^{1/2}. \tag{8.2}$$

Since the body is presumed to be bound gravitationally the minimum radius must also be given by (3.3). Thus, a critical condition is established by the point at which the minimum radius in (3.3) is equal to the minimum holographic radius in (8.2). That condition is

$$\frac{p^2 M_s}{G} \sim \left(\frac{GM_s^{3/2}}{c^2 q^{1/2}}\right)^{1/2}. \tag{8.3}$$

The relationship in (8.3) identifies a fourth critical mass $M_s$ as

$$M_s \sim \frac{G^6}{p^8 c^4 q}. \tag{8.4}$$

The mass in (8.4) is of the order $10^{32}$kg, corresponding to the scale of stellar mass. The cell-radius corresponding to (8.4) is of the order $10^{15}$m, which is the cell-radius of a typical solar system. The minimum radius associated with the mass $M_s$ is, according to (3.3), of the order $10^8$m, which corresponds well to the radius of a typical star.

Any body that is less massive than the critical mass of order near $M_s$ could not be gravitationally bound and part of the self-similar structure of the cosmos if the fractal dimension is 2. Therefore, fractal bodies that are less massive than $M_s$ must be bound by some force other than gravity. That conclusion is important in the context of the analysis in Section 11.

## 9. The nucleon

The minimum possible action in the universe is the Planck constant $\hbar$. The angular momentum and action associated with a fundamental particle must be of the order $\hbar$. In order for a particle, represented by structural level $n$, to be consistent with the physics detailed in Sections 2 – 4, the requirement on its action must be

$$M_n c R_n \sim \hbar. \tag{9.1}$$

The only mass that could satisfy (9.1) and (2.1) is

$$M_n \sim \left(\frac{\hbar^2 q}{c^2}\right)^{1/3}, \tag{9.2}$$

which is roughly $10^{-28}$kg, of order near the nucleon mass $m_n$. The corresponding cell-radius $R_n$ is roughly $10^{-15}$m, corresponding to the Compton wavelength $l_n$ of the nucleon.

Note that the angular momentum law in (3.1) does not apply, *per se*, to the nucleon since $pm_n^2$ is of the order $10^{-70}$Js, which is many orders of magnitude smaller than the Planck quantum. The smallest mass whose angular momentum could be described by (3.1) is $(\hbar/p)^{1/2}$, which is of order near $10^{-9}$kg. That mass is interpreted in Section 11 and it is shown that the scaling law for angular momentum in (3.1) is still physically significant on the scale of particles.

## 10. The cosmic hierarchy and the cosmological constant

Any hierarchy of masses representing the structural levels in a fractal universe should feature a certain signature of self-similarity. Let $N_x$ represent the total number of bodies on some structural level $x$ that are contained within the observable universe. For any two structural levels $j$ and $k$, where $j$ is the immediate parent level of $k$, the numbers $N_j$ and $N_k$ should be related according to

$$N_k \sim N_j^2, \tag{10.1}$$

if the fractal dimension is 2. If $l$ represents the level of structure immediately subordinate to $k$ then $N_l \sim N_k^2 \sim N_j^4$, and so on. Since the total number $N_x$ of observable bodies on any level $x$ is given by $M_0/M_x$, where $M_0$ is the mass of the observable universe, the relationship in (10.1) leads to

$$M_k \sim \left(\frac{M_j}{M_0}\right)^2 M_0. \tag{10.2}$$

It so happens that the masses in (5.2), (6.2), (7.2), (8.4) and (9.2) form naturally a hierarchy that satisfies the condition of self-similarity in (10.2). The hierarchy of mass formed by $M_u$, $M_c$, $M_g$ and $M_s$ is

$$M_u \equiv \frac{c^4}{G^2 q}, \tag{10.3}$$

$$M_c \equiv \frac{c^2}{p^2 q} = \left(\frac{G}{pc}\right)^2 M_u, \tag{10.4}$$

$$M_g \equiv \frac{G^2}{p^4 q} = \left(\frac{G}{pc}\right)^4 M_u, \tag{10.5}$$

and

$$M_s \equiv \frac{G^6}{p^8 c^4 q} = \left(\frac{G}{pc}\right)^8 M_u. \tag{10.6}$$

The self-similarity illustrated in (10.3) – (10.6) is particularly remarkable since the fundamental masses were derived independent of any consideration of the requirement of self-similarity in (10.1) and (10.2).

It must be also that the parameters of the nucleon should be related to the cosmic mass $M_u$, defined in (5.2), according to the self-similar hierarchy in (10.4) – (10.6). However, without some additional physics, the mass $M_n$ in (9.2) does not reduce algebraically to the cosmic mass $M_u$ as do the masses $M_c$, $M_g$ and $M_s$. The mass $M_n$ is nonetheless consistent numerically with the hierarchy, generating significant consequences. According to (10.2) and the hierarchy in (10.3) – (10.6) the next two fundamental masses associated with the fractal universe must be

$$M_{16} \equiv \left(\frac{G}{pc}\right)^{16} M_u, \tag{10.7}$$

and

$$M_{32} \equiv \left(\frac{G}{pc}\right)^{32} M_u. \tag{10.8}$$

The mass $M_{32}$ is of the order $10^{-28}$kg, which is of order near the nucleon mass. The mass $M_{16}$ is of the order $10^{12}$kg and it does not correspond to any of the masses derived here. However, it must have some physical significance in a fractal universe, and that is explored in Section 11.

In addition to validating the expectations of a self-similar hierarchy of mass, the hypothesis that the nucleon mass $m_n \sim M_n$ is physically scaled to $M_{32}$ yields a number of new relationships among the fundamental parameters of nature. With $q$ given by (2.3), it follows from $m_n \sim M_{32}$ that

$$m_n \sim m_P \left(\frac{G}{pc}\right)^8, \tag{10.10}$$

where $m_P \equiv (\hbar c / G)^{1/2}$ is the Planck mass.

While the nucleon establishes a fundamental, microscopic limit on the value of the constant $q$ in (2.1), there may be another fundamental limit on $q$ that is determined by cosmological parameters. The current cosmic mass is of order near the fundamental mass $M_u$ defined in (5.2), and that mass forms the basis of the hierarchy in (10.4) – (10.8). The current cosmic mass is thus special in connection with fundamental parameters and a fractal universe. The fact that the current mass is special becomes particularly significant in the context of the cosmic coincidence and the cosmological constant. The expansion of the universe is apparently accelerating, which may be the result of a cosmological vacuum-energy that is associated with a cosmological constant, $\Lambda$. The existence of a cosmological constant results in fundamental limits on the observable mass and horizon of the universe. The maximum cosmic mass $M_\Lambda$ associated with a cosmological constant is given by [6]

$$M_\Lambda = \frac{c^3}{G\sqrt{\Lambda}}, \tag{10.11}$$

Due to the cosmic coincidence, the current cosmic mass is of order the maximum mass in (10.11). It thus happens that the mass $M_u$ is of order the mass $M_\Lambda$ associated with the putative cosmological constant, and both are of order the current observable mass.

If the observable universe represents the largest structural level of the cosmos and if there is a cosmological constant then the maximum mass $M_u$ in (5.2) must be scaled to the maximum cosmic mass $M_\Lambda$ in (10.11). Consequently, the constant $q$ would be scaled to the cosmological constant according to

$$q \sim \frac{c\sqrt{\Lambda}}{G}. \tag{10.12}$$

The term in (10.12) is of the order $10^0 \text{kg/m}^2$, which is consistent with the value of $q$ determined empirically from the characteristic gravitational fields of astronomical bodies in Section 2. Furthermore, since the term $q$ must be simultaneously given by (10.9) and (10.12), it follows that the nucleon mass must be scaled to the cosmological constant according to

$$m_n \sim \left(\frac{\Lambda \hbar^4}{G^2 c^2}\right)^{1/6}. \tag{10.13}$$

The scaling law in (10.13) has been proposed independently for a variety of different reasons [6],[8],[9],[10],[11]. From (10.10) and (10.13) it follows that the cosmological constant would be given by

$$\Lambda \sim \left(\frac{G}{pc}\right)^{48} \frac{1}{t_P^2}, \tag{10.14}$$

where $t_P$ is the Planck time. The maximum number of degrees of freedom that could be associated with our universe is roughly $c^5/(G\hbar\Lambda)$ [6], and it would be given by

$$\frac{c^5}{G\hbar\Lambda} \sim \left(\frac{G}{pc}\right)^{-48}. \tag{10.15}$$

*11. Completing the hierarchy*

The term $M_{16} \sim 10^{12}$kg in (10.7) represents necessarily a structural level of the fractal hierarchy, but its significance has not been established from first principles as was done for the other masses. Note that any body with a mass less than $M_s$ may not be both gravitationally bound and part of the fractal structure of the universe with dimension 2. Thus, any physical significance to $M_{16}$ must involve some other force of nature. Given that the nuclear force has already been associated with $M_{32}$, it is reasonable to suspect that the mass $M_{16}$ is associated with the electrostatic force. Such a connection is readily identified. The electrostatic force between a proton and electron is equal to the gravitational force between a mass of the order $M_{16}$ and an electron.

If the mass $M_{16}$ does represent the gravitational equivalent of the electrostatic force of the proton on the electron then the fundamental charge $e$ and the mass $m_e$ of the electron would be scaled to $M_{16}$ according to

$$k_e e^2 \sim GM_{16} m_e, \quad (11.1)$$

where $k_e$ is the electrostatic force constant in SI units. If (11.1) represents a physical relationship then it follows from (10.10), (11.1) and that fact that $M_{16}/M_{32}$ is equal to $(G/(pc))^{-16}$ that the fine-structure constant $\alpha \equiv k_e e^2/(\hbar c)$ would be scaled according to

$$\alpha \sim \frac{m_e}{m_n}. \quad (11.2)$$

Moreover, it would follow from (10.10), (11.1) and (11.2) that

$$\alpha \sim \frac{G}{pc}. \quad (11.3)$$

Wesson has noted that the similarity between $\alpha$ and the ratio $G/(pc)$ may be significant, and the hypothesis concerning $M_{16}$ would validate some well-justified suspicions about the fundamental constants [5].

It is also compelling that the hypothesized scaling in (11.3) would allow the scaling law for angular momentum in (3.1) to be extended to the particle scale. The effective gravitational mass of the fundamental charge unit is $m_k \equiv (k_e e^2/G)^{1/2}$. (The mass $m_k$ is the geometrical mean of $M_{16}$ and $m_e$.) It follows from (11.3) that the mass $m_k$ is roughly $(\hbar/p)^{1/2}$. That mass is significant since the angular momentum associated with the mass $(\hbar/p)^{1/2}$ is, according to (3.1), the Planck quantum $\hbar$. Thus, the relevance of the scaling law in (3.1) is recovered for particles whose angular momentum is the Planck quantum if, instead of the particle mass, the effective mass $m_k$ of the fundamental charge is used in the scaling law, since it would be that $pm_k^2 \sim \hbar$.

It is interesting to note that the mass $l_n c^2/G$ of the black hole whose horizon is the Compton wavelength $l_n$ of the nucleon is of the order $10^{12}$kg. It follows from the proposed scaling law in (10.10) that the mass $M_{16}$ would necessarily be $\sim l_n c^2/G$. Thus, according to (11.1), the gravitational equivalent of the electrostatic force of the proton on the electron would be scaled to the mass of a black hole whose horizon is the Compton wavelength of the proton.